%% file: jharris.tex
\documentclass[12pt,preprint]{aastex}
\begin{document}
\def\eg{{\it e.g.}}
\def\ie{{\it i.e.}}
\newcommand{\tnm}[1]{\tablenotemark{#1}}
\newbox\grsign
\setbox\grsign=\hbox{$>$}
\newdimen\grdimen
\grdimen=\ht\grsign
\newbox\simlessbox
\newbox\simgreatbox
\setbox\simgreatbox=\hbox{\raise.5ex\hbox{$>$}\llap
     {\lower.5ex\hbox{$\sim$}}}\ht1=\grdimen\dp1=0pt
\setbox\simlessbox=\hbox{\raise.5ex\hbox{$<$}\llap
     {\lower.5ex\hbox{$\sim$}}}\ht2=\grdimen\dp2=0pt
\def\simgreat{\mathrel{\copy\simgreatbox}}
\def\simless{\mathrel{\copy\simlessbox}}

\title{The Star Formation History of the Small Magellanic Cloud}

\author{Jason Harris\altaffilmark{1}}
\affil{Space Telescope Science Institute}
\affil{3700 San Martin Dr., Baltimore, MD, 21218}
\affil{E-Mail: jharris@stsci.edu}
\author{Dennis Zaritsky}
\affil{Steward Observatory}
\affil{933 North Cherry Ave., Tucson, AZ, 85721}
\affil{E-Mail: dzaritsky@as.arizona.edu}

\altaffiltext{1}{current address: Steward Observatory, 933 N.
Cherry Ave., Tucson, AZ, 85721}

\begin{abstract}
We present the spatially-resolved star formation and chemical
enrichment history of the Small Magellanic Cloud (SMC) across the
entire central $4^\circ \times 4.5^\circ$ area of the main body,
based on $UBVI$ photometry from our Magellanic Clouds Photometric
Survey.  We find that 1) approximately 50\% of the stars that ever
formed in the SMC formed prior to 8.4~Gyr ago ($z >1.2$ for WMAP
cosmology), 2) the SMC formed relatively few stars between
8.4 and 3~Gyr ago, 3) there was a rise in the mean star formation rate
during the most recent 3 Gyr punctuated by ``bursts'' at ages of
2.5, 0.4, and 0.06~Gyr, 4) the bursts at 2.5 and 0.4~Gyr are
temporally coincident with past perigalactic passages of the SMC with
the Milky Way, 5) there is preliminary evidence for a large-scale
annular structure in the 2.5~Gyr burst, and 6) the chemical enrichment
history derived from our analysis is in agreement with the
age-metallicity relation of the SMC's star clusters.  Consistent
interpretation of the data required an ad hoc correction of
0.1--0.2~mag to the B-V colors of 25\% of the stars; the cause of this
anomaly is unknown, but we show that it does not strongly influence our
results.
\end{abstract}

\keywords{ galaxies: evolution ---
galaxies: stellar content ---
galaxies: Magellanic Clouds ---
galaxies: individual: Small Magellanic Cloud }

\section{Introduction}\label{sec:intro}

Determinations of detailed, quantitative star formation histories
(SFHs) of local galaxies aim to provide an empirical foundation
upon which a comprehensive theory of star formation in galaxies can
be constructed.  Presently, even basic questions of how star
formation proceeds on galactic scales remain unanswered.  Do
galaxies form stars continuously, or in bursts separated by epochs
of relative quiescence?  If star formation occurs in bursts, what
processes mediate the bursts?  Major galaxy interactions are known
to induce vigorous star-formation events \citep[\eg,\ ][]{lt78,
lon84, cm85}, but do less-dramatic interaction events also trigger
significant star formation?  Are parametrizations describing
stellar populations, such as the initial mass function (IMF) or the
cluster-to-field star ratio, determined by the integrated
dynamical interaction history of a galaxy, or are they universal?
How significant are gas inflow and outflow to the chemical enrichment
history of galaxies?

These long-standing questions can be addressed directly by studying
the detailed SFHs of Local Group galaxies.  The Magellanic Clouds
are optimal targets for SFH analysis because 1) their proximity
allows us to obtain color-magnitude diagrams (CMDs) of resolved
stellar populations well down the main sequence and measure a bulk
proper motion from which their space velocity and orbit can be
derived, 2) their proximity to the Milky Way (and to each other)
suggests that tidal interactions may play an important, periodic role
in triggering star formation, and 3) their on-going star formation 
enables us to measure star formation events at recent times, where
we can achieve the high temporal resolution necessary to investigate 
possible triggering mechanisms and the the high spatial resolution
necessary to investigate the interplay between star formation and the 
interstellar medium.

Previous studies of the Clouds have resulted in tantalizing glimpses
of their  SFHs.  For example, the ``age-gap" among LMC clusters
\citep{vdb91, dc91, gir95, wes97} suggests that there was a long 
period of quiescence in its history.  In a photographic-plate study
of the outer regions of the SMC, \cite{gh92} found that the bulk of
the stellar population is about 10~Gyr old, with about 7\% of the 
population aged 15--16~Gyr, and also observed a young stellar
population biased toward the eastern, LMC-facing side of the SMC.  
\cite{cro01} see a similar trend among the SMC's populous clusters: 
those on the eastern side tend to be younger and more metal-rich
than those on the western side.  Finally, several authors have 
performed SFH analyses on HST/WFPC2 fields in both Clouds, offering 
detailed snapshots of the histories of specific regions in 
these galaxies \citep{glg96, ard97, hol99, ols99, dol01, sh02}.
However, there has not yet been an attempt to determine the full, 
global star-formation history of either Magellanic Cloud, because a 
sufficiently sensitive, spatially comprehensive catalog of
their stellar populations had not been available prior to our
Magellanic Clouds Photometric Survey (MCPS; \cite{zar97}).

In this article, we present the global star-formation history of the
Small Magellanic Cloud, based on the MCPS $UBVI$ catalog that
includes over six million SMC stars. We briefly review the MCPS and
the StarFISH star-formation history reconstruction program
\cite{hz01} in Section~\ref{sec:mcps}.  The determination of the
necessary inputs to StarFISH (including detailed treatments of the
interstellar extinction and photometric errors) is presented in
Section~\ref{sec:inputs}.  We describe the application
of StarFISH to the SMC catalog in Section~\ref{sec:starfish}, and
our comprehensive map of the SMC's SFH is presented in
Section~\ref{sec:map}.  Analysis and discussion of the SFH map,
which shows significant structure in both the spatial and temporal
dimensions, is presented in Section~\ref{sec:discuss}, and we
summarize our findings in Section~\ref{sec:summary}.

\section{Deriving the SMC's Star Formation History}\label{sec:derive}

\subsection{A Brief Summary of the Previously Published Data and
Methodology}\label{sec:mcps}

\begin{figure}[h]
\plotone{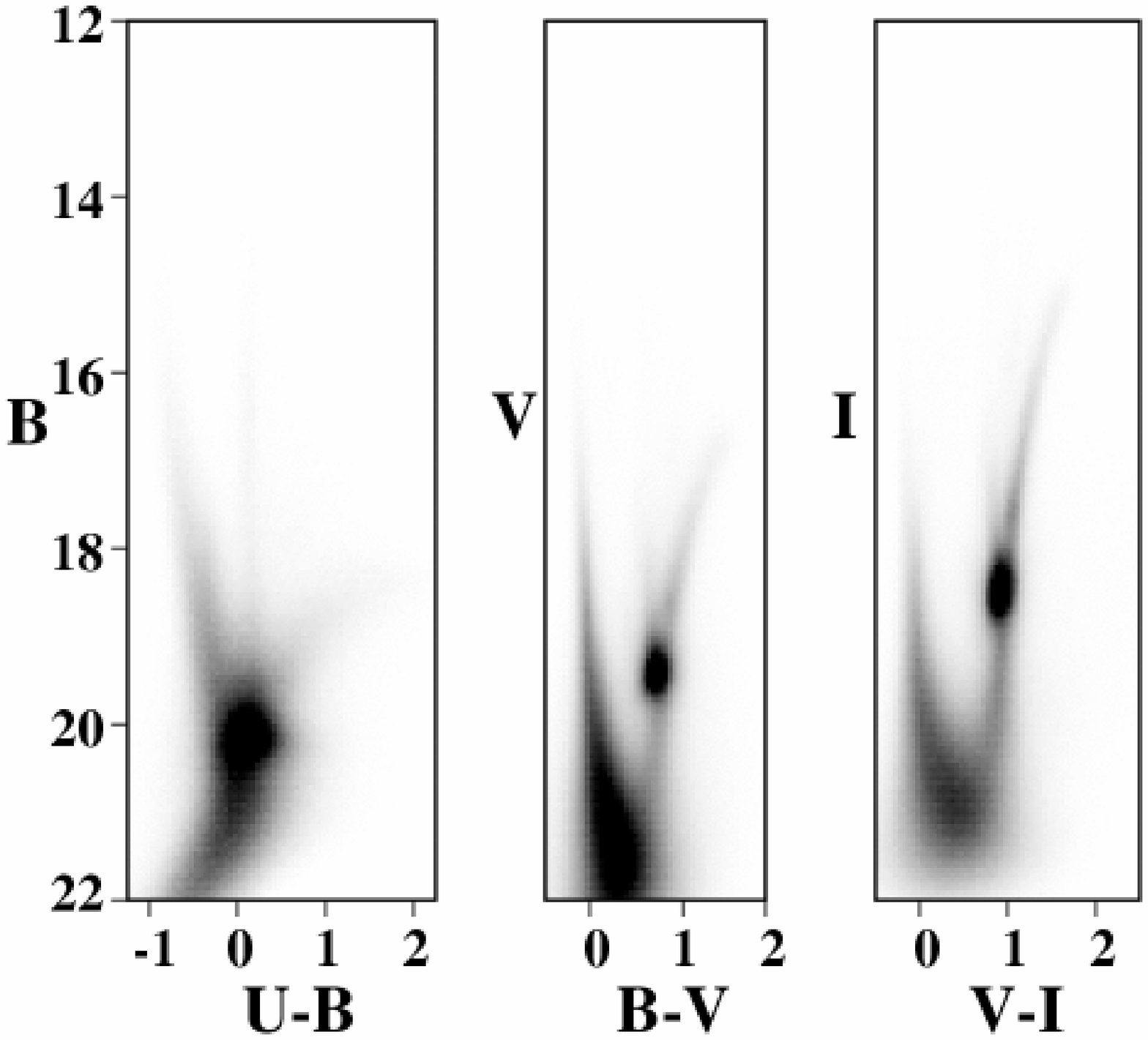}
\caption{A ``CMD triptych'' illustrating our $UBVI$ photometry of
six million SMC stars.  Each CMD panel is a pixelized Hess diagram
showing the number of stars in each pixel.  The mixed populations
evident in these CMDs represent a ``fossil record'' of a complex
star-formation history in this galaxy. \label{fig:ubvi} }
\end{figure}

Our Magellanic Clouds Photometric Survey \citep[MCPS,\ ][]{zar97}
provides the most complete optical survey of bright stellar 
populations in the Magellanic Clouds to date.  The Survey was 
conducted between 1996 November and 1999 December at the Las 
Campanas Observatory 1-meter Swope Telescope.  We employed
a unique drift-scan CCD instrument \citep[the Great Circle
Camera,][]{zar96} which allowed us to efficiently acquire large,
distortion-free scans at the extreme declinations of the Magellanic
Clouds.  Typical survey scans are 24$\arcmin$ wide and 2$\arcdeg$
long. Twenty-six scans cover the $4\arcdeg\times4.5\arcdeg$ area
surveyed in the SMC.  Each scan was observed using $U$, $B$, $V$,
and $I$ filters and the effective exposure times are $\sim$5
minutes.  We use an automated data-reduction pipeline that consists
of IRAF \citep{iraf86} scripts and the DAOPHOT photometry package
\citep{ste87}.  The final SMC catalog \citep{zar02} is 50\% complete
to $m_v=21--22$~mag, depending on the local crowding conditions.
The catalog includes at least $B$ and $V$ photometry for over
6~million SMC stars (see Figure~\ref{fig:ubvi}). In the present
analysis, we use a subset of the photometry with $V \le 21$, where
the completeness corrections are modest (50\%--95\% depending on local
crowding conditions).

We developed the StarFISH package \citep{hz01} to determine the
detailed star-formation histories encoded in the stellar
populations of our SMC and LMC catalogs (however, the package is
designed to be generally applicable to any photometric data and is
available for public use).  StarFISH performs a chi-squared
minimization between the observed photometry and model photometry
based on theoretical isochrones (we employ the Padua isochrones
most recently published by \cite{gir02}, but the package is
sufficiently flexible to use of any set of isochrones).  We
construct a library consisting of sets of three synthetic Hess
diagrams, which we shall refer to as {\it CMD triptychs}.  Each CMD
triptych represents the predicted $UBVI$ photometry for a single
stellar population of a specific age and metallicity.  Constructing
the synthetic CMD triptychs requires us to specify the
distance, initial mass function, binary fraction, and the
empirical models of the interstellar extinction and photometric
errors.  Once we have the library of synthetic CMD triptychs, we
construct a {\it composite model CMD triptych} that represents the
predicted photometry for any arbitrary SFH.  The composite model is a
linear combination of the set of synthetic CMD triptychs, each of
which is modulated by an amplitude value that is equal to the number
of stars present at the age and metallicity of the corresponding
synthetic population.  We employ an efficient downhill simplex
algorithm to select the model triptych that is most similar to that
observed and evaluate uncertainties by examining the parameter space
about the best fit set of amplitudes \citep[see][]{hz01}.
Although the fitting could be done purely in the multidimensional
color space,
the use of the triptych is adopted for visualizing the fits and potential
systematic errors.

\subsection{Defining the Inputs to StarFISH}\label{sec:inputs}

As described broadly above, a number of parameter values must be
set before running StarFISH. We adopt a distance modulus
for the SMC of 18.9~mag \citep{dol01}, corresponding to
$\sim60$~kpc.  For the IMF, we adopt a simple power-law with a
Salpeter slope \citep{sir02}.  Having little solid information to
guide us in adopting a binary fraction, we simply adopt a fraction
of 0.5, with secondary masses drawn randomly from the IMF
\citep[consistent with the findings of ][]{dm91}.  Our detailed,
statistical modeling of interstellar extinction and photometric
errors are discussed below.

\subsubsection{The Isochrone Set}\label{sec:isoc}

For the present analysis, we begin with a subset of the Padua
isochrones \citep{gir02} for three metallicities appropriate for
stellar populations in the SMC: Z=0.001, 0.004, and 0.008
(corresponding to [Fe/H]=-1.3, -0.7, and -0.4).  We note that
this may be a bit more metal-rich than the most metal-poor
populations in the SMC.  Notably, NGC~121 (the
SMC's oldest populous cluster) has [Fe/H] between -1.7 and -1.0
\citep{sun86, msf98, dh98, dfp98, dol01}.  The chemical
enrichment model for the SMC by \cite{pt99} indicates rapid metal
enrichment in its early history, rising above [Fe/H]=-1.3
around 11~Gyr ago (see their Figure~5).  We therefore believe
that the omission of more metal-poor isochrones in our analysis
will not significantly impact our results, since any stellar
populations with metallicity below Z=0.001 are expected only in
our very oldest age bin.

We deemed it unecessary to interpolate between these three
metallicity bins.  Our tests show that StarFISH
can account for intermediate metallicities by simply mixing
amplitudes of the bounding isochrones \citep[see analysis of NGC~1978
in ][]{hz01}; in other words, there are no
significant ``gaps'' between isochrones of the same age and adjacent
metallicities in our synthetic CMD triptychs.  This conclusion is
data-dependent, of course; higher-precision photometry than the MCPS
may well require finer metallicity resolution than we use here.

For each metallicity, the Padua group provides isochrones for 62 ages
distributed uniformly in log(age) between 4~Myr and 18~Gyr.  This
time resolution is much too fine for our purposes; when interstellar
extinction and photometric errors are included, some isochrones with
adjacent ages are completely degenerate.  As outlined in \cite{hz01},
we circumvent this degeneracy by ``locking'' together isochrones into
groups of four, yielding an effective age resolution of 
$log(\Delta age)=0.2$.  Thus, each of our synthetic CMD triptychs
actually spans a range of ages, and the age bins are wide enough that 
no CMD triptych is completely degenerate with any other.  Again, this 
age resolution was adopted empirically, taking the photometric error
characteristics of the data into account.  Our adopted synthetic CMD
library consists of 47 isochrone groups; 11 ages spanning 100~Myr to
12~Gyr for Z=0.001, and 18 ages spanning 4~Myr to 12~Gyr for both
Z=0.004 and Z=0.008.

\subsubsection{Interstellar Extinction}\label{sec:dust}

In a previous analysis of the interstellar extinction in the SMC
\citep{zar99}, we found (a) the distribution of extinction values is
much wider than can be explained by photometric errors; \ie, there is
significant intrinsic differential extinction in the SMC, and (b)
interstellar extinction exhibits a strong dependence on stellar
population type: the extinction values toward hot stars ($T >
12000$ K) are typically four times larger than those toward cooler
stars ($T \sim 6000$ K). This result is not surprising given the
dustier environments in which younger stellar populations are
generally found, but we are unaware of another synthetic-CMD
technique that accounts for such population-dependent extinction.

\begin{figure}[h]
\plotone{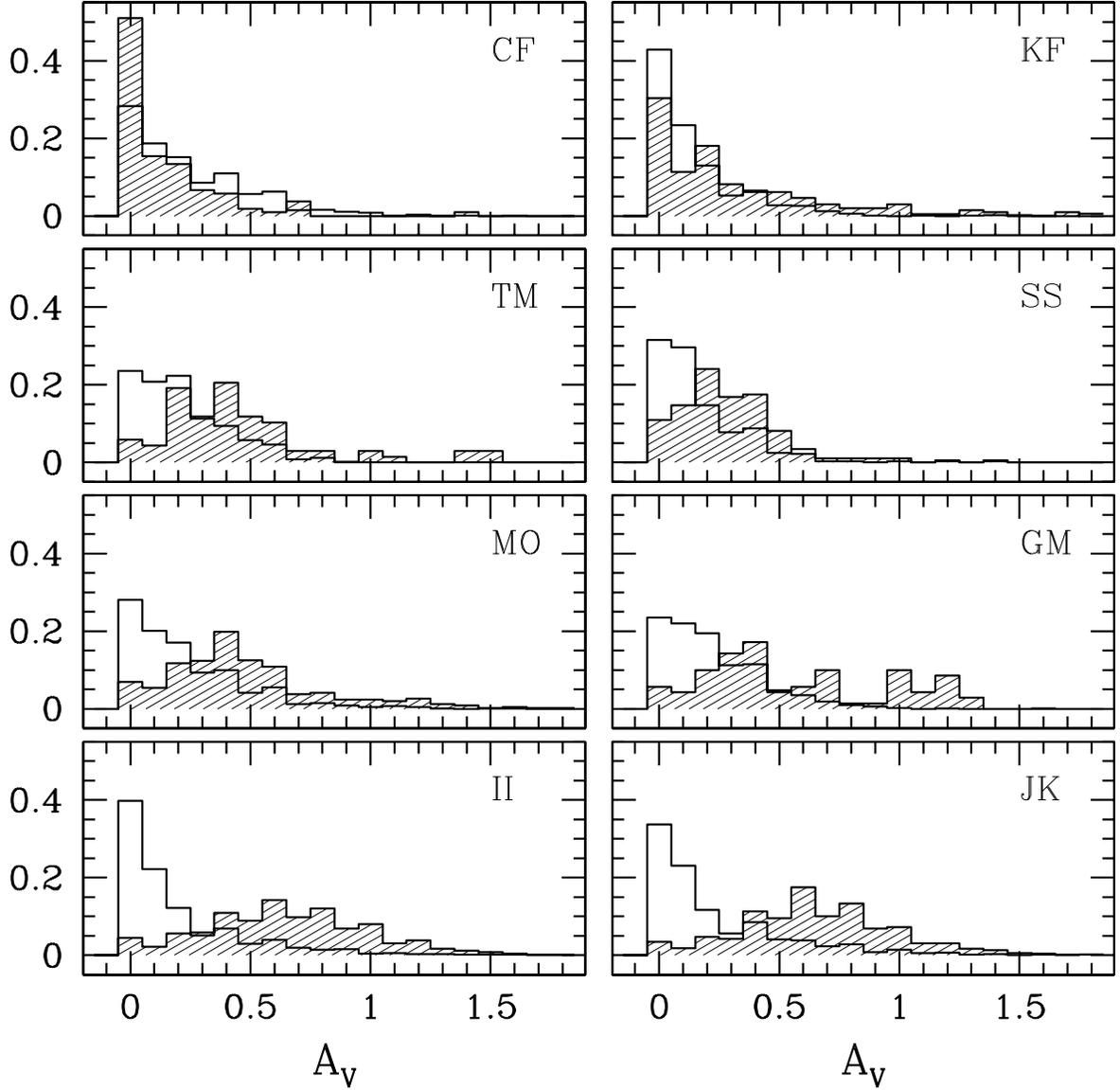}
\caption{The measured extinction distributions from eight selected
SMC subregions.  Each panel shows both the cool-star sample (open
histogram) and the hot-star sample (shaded histogram) for stellar
populations drawn from that particular SMC subregion.  The subregion
is identified by the two-letter code in the top-right corner of
each panel (see Figure~\ref{fig:grid}).  The subregions were
selected to illustrate the spatial variation in the extinction
properties of the SMC.  While the cool-star extinction distributions
are rather uniform throughout the SMC, the hot-star extinction
distributions vary substantially.  In addition, the hot-star and
cool-star distributions are always much wider than the measurement
errors, making it inadvisable to characterize the extinction in any
region of the SMC with a single value. \label{fig:extdist} }
\end{figure}

Using our extinction measurements for stars in each of these two
temperature regimes, we construct hot- and cool-star extinction
distributions for each of our 351 SMC regions (see
Figure~\ref{fig:extdist}).  We adopt the hot-star extinction
distribution to create synthetic CMDs for populations with ages
younger than 10~Myr, and the cool-star extinction distribution for
populations with ages older than 1~Gyr.  For populations with ages
between 10~Myr and 1~Gyr, we adopt a linear combination of the two
distributions, with a statistical weight that linearly favors the
cool-star distribution as log(age) increases and matches the limiting
values at the two bounding ages.  When generating model stars for a
region's synthetic CMDs, extinction values are drawn randomly from
the measured empirical distributions for the region. In this manner
we reproduce both the observed correlation between
population age and extinction, and the intrinsic differential
extinction appropriately for each individual region.

\subsubsection{Photometric Errors}\label{sec:photerrs}

Many factors contribute to photometric errors: seeing, atmospheric
transparency, variable sky levels, crowding, CCD readnoise, and in
the case of the MCPS, a variable PSF that may result when the GCC's
drift-scanning paramters are not ideal.  Many of these contributors
can exert a position-dependent effect on the catalog photometry,
especially because the data were obtained on dozens of nights under
a variety of conditions, spread over four years of observations.
Therefore, the photometric error characteristics of each subregion
would ideally be modeled using artificial star tests (ASTs) performed
on that particular subregion's images.  However, the computational time required to
perform hundreds of ASTs, each composed of hundreds of thousands of
artificial stars added in dozens of trials to each image, is
currently prohibitively large.  By examining a variety of images from
the survey, we determine that of all the observational effects
listed above, the effective photometric errors in the MCPS are
generally dominated by crowding effects.  For two typical SMC
subregions of similar stellar surface density, their photometric
error characteristics are statistically indistinguishable.  Using
this characteristic of the survey, we greatly reduce the number of
required artificial stars tests by applying the results of one set of
ASTs to all images of similar stellar surface density.  Our adopted
strategy is to select eight subregions to provide representative AST
results that span the range of stellar surface densities in the SMC.
Table~\ref{tab:ast} lists the regions for which we have performed
ASTs and their stellar densities.

To perform the ASTs, we use the DAOPHOT ADDSTAR program to add
artificial stars to the selected images using the point-spread
function (PSF) determined from the stars in  the subregion.  The
artificial stars are assigned right ascension and declination (RA,
Dec) coordinates in a fixed grid throughout the image spaced by 20
pixels in each direction, so that the artificial stellar profiles
overlap only beyond their their $5\sigma$-equivalent radii, ensuring
that they sample the crowding environment without affecting it by
their presence (\ie, each artificial star's profile is guaranteed to
be affected only by real objects, not other artificial stars). This
grid strategy limits the number of artificial stars that can be added
to a single image to several hundred.  Therefore, we perform a large
number of such trials to accumulate a large sample of artificial
stars.  In each trial, the zeropoint of the coordinate grid is given
a random offset so that each trial's artificial stars sample new
crowding conditions in the frame.  When adding the artificial stars
to each of the $U$, $B$, $V$ and $I$ images, we invert the frame's
coordinate transformation to obtain X,Y pixel coordinates for the
artificial stars from their original RA, Dec coordinates so that
the stars are coincident on the sky, rather than on the CCD, in the
various filters.  For strongly clustered stellar populations (star
clusters or very young stars \citep{hz99}), this procedure
underestimates the effect of crowding, but these are two minor
components of the entire stellar population of either the SMC or
LMC.

To determine photometric uncertainties and completeness fractions, we
analyze each new image using the same data-reduction pipeline used to
determine the original photometry (with the sole exception that
instead of solving for the PSF we adopt the original best-fit PSF).
This procedure results in three photometric catalogs for each
artificial-star image: the {\it original catalog}, which contains
only real stellar photometry,  the {\it intrinsic AST catalog} which
contains the input photometry for the artificial stars,  and the
{\it observed AST catalog}, which contains photometry for both real
stars and artificial stars.  We need to match each star in the
intrinsic AST catalog to its corresponding detection in the observed
AST catalog.  This task is complicated because the star fields are
typically quite crowded.  To reduce confusion, we first match stars
in the original catalog to stars in the observed AST catalog,
retaining only those objects in the observed AST catalog which are
{\it not} matched to objects in the original catalog.  To minimize
the matching of artificial stars (or real-artificial blends) to real
stars, we impose a small matching radius (0.5 pixels) and also
require that the photometry is the same to within $\pm0.5$~mag in
each filter between the original and observed AST catalogs. The
unmatched objects are then matched to the intrinsic AST catalog to
produce a list of input and output photometry for the artificial
stars.  If no match could be found for an artificial star in the
observed catalog, the artificial star is flagged as a dropout in that
image.  StarFISH constructs both a photometric error model
and the completeness rate as a function of position in the CMDs
directly from the table of input and recovered photometry of
artificial stars.

\subsection{Running StarFISH}\label{sec:starfish}

\subsubsection{Spatial Partitioning}\label{sec:grid}

\begin{figure}[h]
\plotone{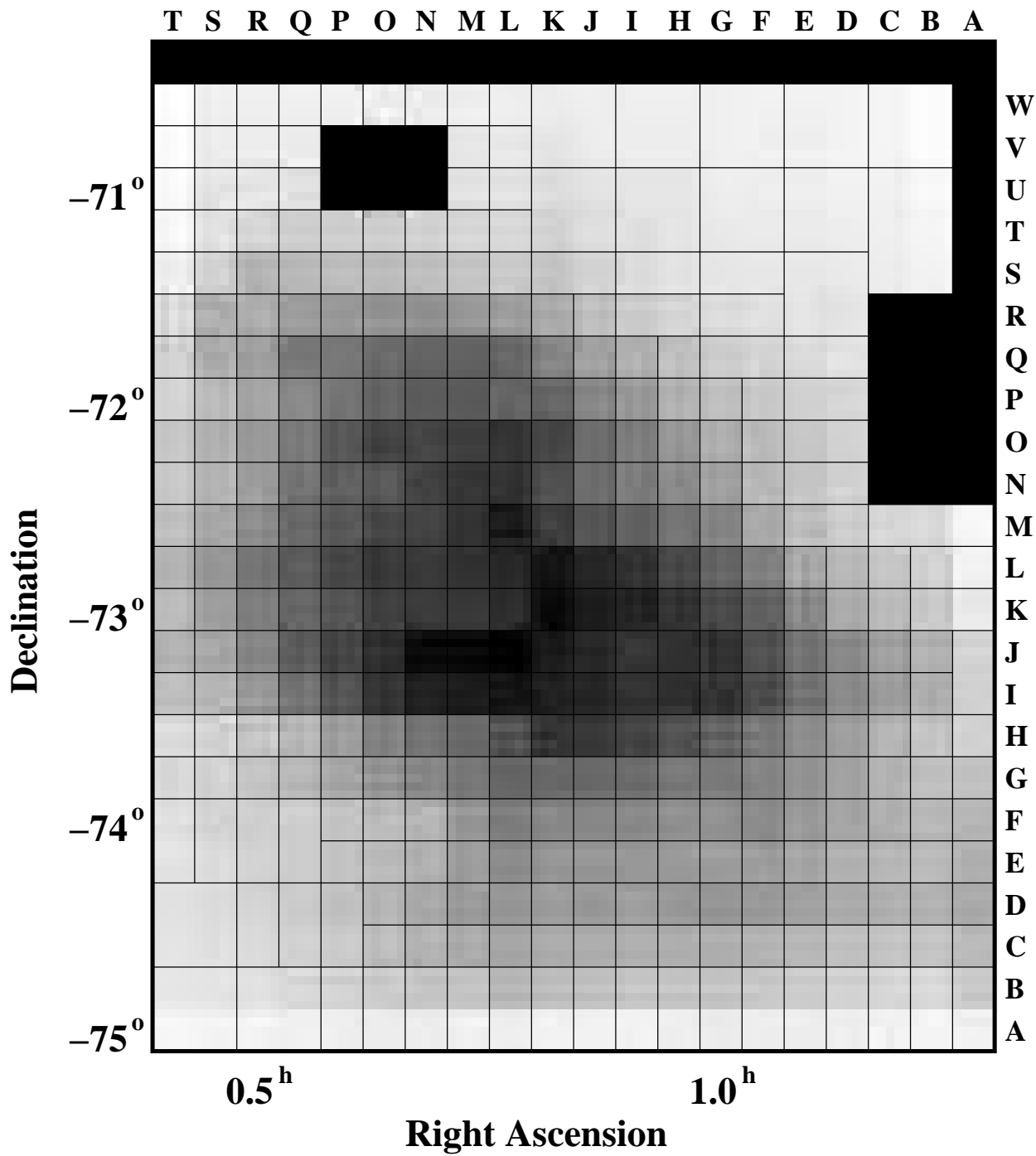}
\caption{The division of our SMC catalog into 351 subregions.  The
greyscale image shows the number of stars present in our MCPS catalog
from each subregion (where white means zero stars, and black means
approximately 30,000 stars).  The primary division imposes a uniform
$20\times23$ grid of subregions, each approximately
$12\arcmin\times12\arcmin$ in extent.  Where the density of stars is
very low, we combine adjacent grid cells into larger subregions.
We mask some regions where foreground contamination (due
to Galactic globular clusters along the line of sight) is significant.
The large masked region on the west edge (regions AN through CR) is
due to 47~Tucanae; the smaller masked region near the North edge is
due to NGC~362. \label{fig:grid} }
\end{figure}

To derive a spatially-resolved SFH, we divide the SMC Survey using a
rectilinear grid of 351 subregions, as shown in
Figure~\ref{fig:grid}.  We use a two-letter code to identify
subregions by their position in the grid. The first letter identifies
a subregion's Right Ascension position in the grid, while the second 
letter identifies its Declination position.  The letters are in
alphabetical order for the direction of increasing RA or Dec.  The angular size of
the grid cells is a compromise between wanting small cells for finer
spatial resolution in the map, and needing a large number of stars in
each region for the StarFISH analysis.  Some of the sparsely-
populated grid cells in the outer parts of the survey region were
combined into larger subregions because they did not individually
contain a sufficient number of stars.  Testing shows that we
minimally require of order $10^4$ stars for a stable SFH solution.
A small number of cells have been masked out and are not modeled (see
Figure~ \ref{fig:grid}.  These cells are contaminated by foreground
Galactic globular clusters (the masked region near the western edge
of the survey region is due to 47~Tucanae; the smaller region near the
northern edge is due to NGC~362).

In addition to the goal of creating a spatially-resolved SFH map
for the SMC, the division of our catalog into small regions was
necessary from a practical standpoint as well.  Quantitative
CMD-fitting algorithms such as StarFISH depend sensitively on
accurate statistical representations of both interstellar
extinction and photometric errors.  Since our photometric catalog
covers the entirety of the SMC, it includes a wide variety of
extinction and crowding conditions, making it impossible to apply
these effects in a uniform way for the entire catalog.  By
subdividing the catalog into small regions, we account for
the extinction and photometric errors locally and independently
for each region, which greatly improves the correspondence between
the observed photometry and the model CMD triptychs.

\subsubsection{Finding the Best-Fit Model}\label{sec:chi}

For each of our 351 subregions, we have the $UBVI$ photometry and a
library of synthetic CMD triptychs that incorporate the derived
photometric errors and extinction distributions directly from the
data.  Given these inputs, StarFISH determines the set of amplitude
values modulating each synthetic CMD triptych that produces the best
fit between the observed photometry and the composite model CMD
triptych by using a downhill simplex algorithm to evaluate the
$\chi^2$ statistic of different composite models \citep[see][]{hz01}.
To evaluate $\chi^2$, the observed and model CMD triptychs are
divided onto a uniform grid with cells that are 0.25~mag wide in both
the color and magnitude directions. The number of observed and model
stars present in each grid cell are then compared using the standard
$\chi^2$ formula:

$$\chi^2 = \sum_i \frac{( N_d(i) - N_m(i) )}{N_d(i)}$$

\noindent
where $N_d(i)$ is the number of stars observed in CMD region $i$,
and $N_m(i)$ is the number of stars in the composite model in CMD
region $i$.  In cases where $N_d(i)$ is zero (but $N_m(i)$ is not),
the denominator is instead taken to be 1.  Regions where both
$N_d(i)$ and $N_m(i)$ are zero do not contribute to $\chi^2$.
Technically, $\chi^2$ should only be used in the case of normally-
distributed errors, which is only approximately true in our case when
both $N_d(i)$ and $N_m(i)$ are large. The $\chi^2$ minimum
can still be used to determine the best-fit model in the case of
non-Gaussian errors, but one loses some ability to determine the
quality of the best fit and the confidence intervals about that
best-fit.  Our testing of the StarFISH algorithm shows that
$\chi_\nu^2\sim1$ does provides a reasonable indication of a good
fit \citep[see\ ][]{hz01} for the number of stars in a typical grid
cell.

\begin{figure}[h]
\plotone{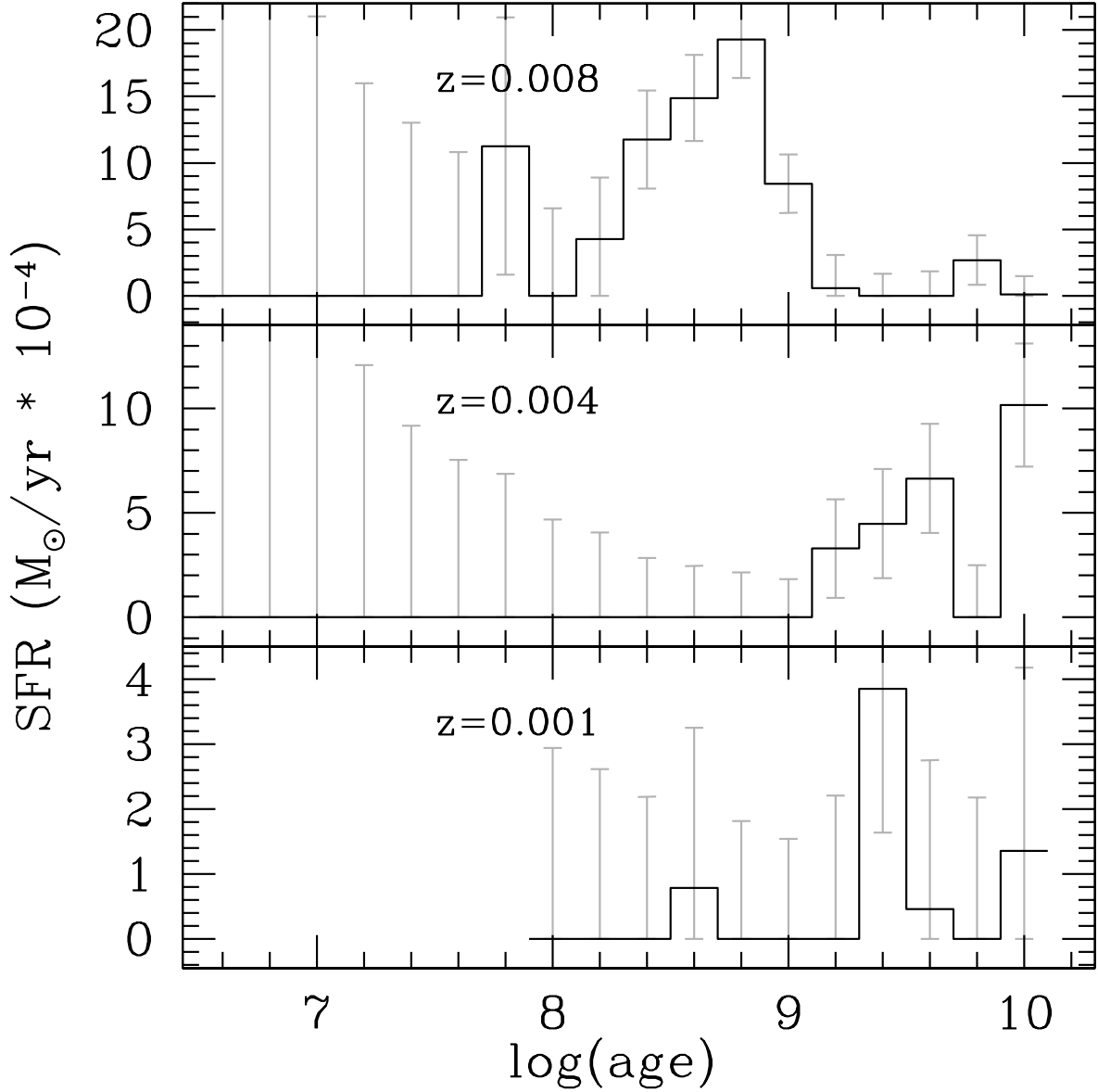}
\caption{The SFH solution for one of our 351 regions in the SMC
(the region labeled MK in Figure~\ref{fig:grid}).  The histogram
indicates the best-fit star formation rates for 47 logarithmic age
bins, across three metallicities: Z=0.008 (top panel), Z=0.004 (middle
panel), and Z=0.001 (bottom panel).  The errorbars represent the
1-$\sigma$ confidence interval on each amplitude, including covariance
between amplitudes. \label{fig:sfhMK} }
\end{figure}

In Figure~\ref{fig:sfhMK}, we show the SFH solution for a typical
region from our SMC grid (region MK).  This region contained
25,848 stars, and the SFH fit solution has a reduced $\chi^2$
value of 3.

\subsubsection{Estimating Fit Uncertainties}\label{sec:fiterrs}

Once the best-fit amplitudes have been determined, the program
performs a systematic exploration of the parameter space surrounding
the best-fit point to determine the $1\sigma$ confidence interval on
each parameter value as defined by the appropriate value of
$\Delta\chi^2 = \chi^2 - \chi_{min}^2$.
The exploration is performed in stages.  First, each amplitude value
is varied while all others are held fixed at their best-fit values.
This calculation evaluates the independent uncertainty associated
with each amplitude value.  Second, adjacent pairs of amplitudes are
varied simultaneously, while the remaining amplitudes are held fixed
at their best-fit values.  This evaluates the correlated errors
between the adjacent amplitude pairs.  Third, we perform a
correlated-error analysis involving the variation of all amplitude
values simultaneously.  We select a random ``direction" in the 47-
dimensional parameter space and evaluate $\chi^2$ for points
displaced from the best-fit point along that direction.  We continue
stepping away from the best-fit point until the $\Delta\chi^2$ value
indicates that we have reached the $1\sigma$ confidence interval.
The $\Delta\chi^2$ evaluation is repeated for 30,000 different random
parameter space directions.  We performed tests of the growth of the
confidence intervals as the number of random directions evaluated
increases.  These tests show steady growth of the confidence intervals
up to about 10,000 directions, and little further growth thereafter.
We conclude that exploring 30,000 directions is sufficient to provide
a robust determination of the correlated errors on the amplitude
values.  Note that without having run stages one and two first, the
number of directions required for the third stage to converge would
be much larger.  This is because we expect each amplitude to have an
independent uncertainty, and we also expect strong correlated errors
between adjacent amplitudes.  Relying on a random selection of
directions in a 47-dimensional space to cover these particular
directions would be extremely inefficient.  Instead, we manually
explore those directions where we expect large deviations, and
use the random-direction stage to fill in any unexpected
correlations.

Throughout each of the three confidence-interval stages (uncorrelated
errors, pairwise correlated errors, and full correlated errors), the
program keeps track of the maximum variation of each amplitude value
that resulted in a $\Delta\chi^2$ value within the confidence
interval.  The final maximum variation defines the endpoints of the
$1\sigma$ confidence interval assigned to each amplitude (see Figure~\ref{fig:sfhMK} for a typical example of the SFH errors).  Note
that because correlated errors between amplitudes are important, the
confidence-limit ``error bars'' are larger than the random
uncertainty of the SFH solution. It can therefore be misleading to
judge the significance of a fluctuation in the SFH relative to its
neighboring amplitudes simply by comparing the difference in star
formation rates to the plotted error bars.

\subsubsection{Problem Areas}\label{sec:bad}

After a first-pass run of StarFISH on all 351 regions, we found that
the best-fit reduced $\chi_\nu^2$ values in some regions were greater
than 10. These poorly-fit regions were in the extremely crowded
central parts of the SMC (see Figure~\ref{fig:grid}). By adjusting
which AST region we used to generate the synthetic CMD library (see
Table~\ref{tab:ast}), we could often improve the fit sufficiently to
bring $\chi_\nu^2$ below 10.  However, the need to do this manual
adjustment indicates that our hypothesis that stellar surface density
dominates the photometric errors does not hold true in every case.

\begin{figure}[h]
\plotone{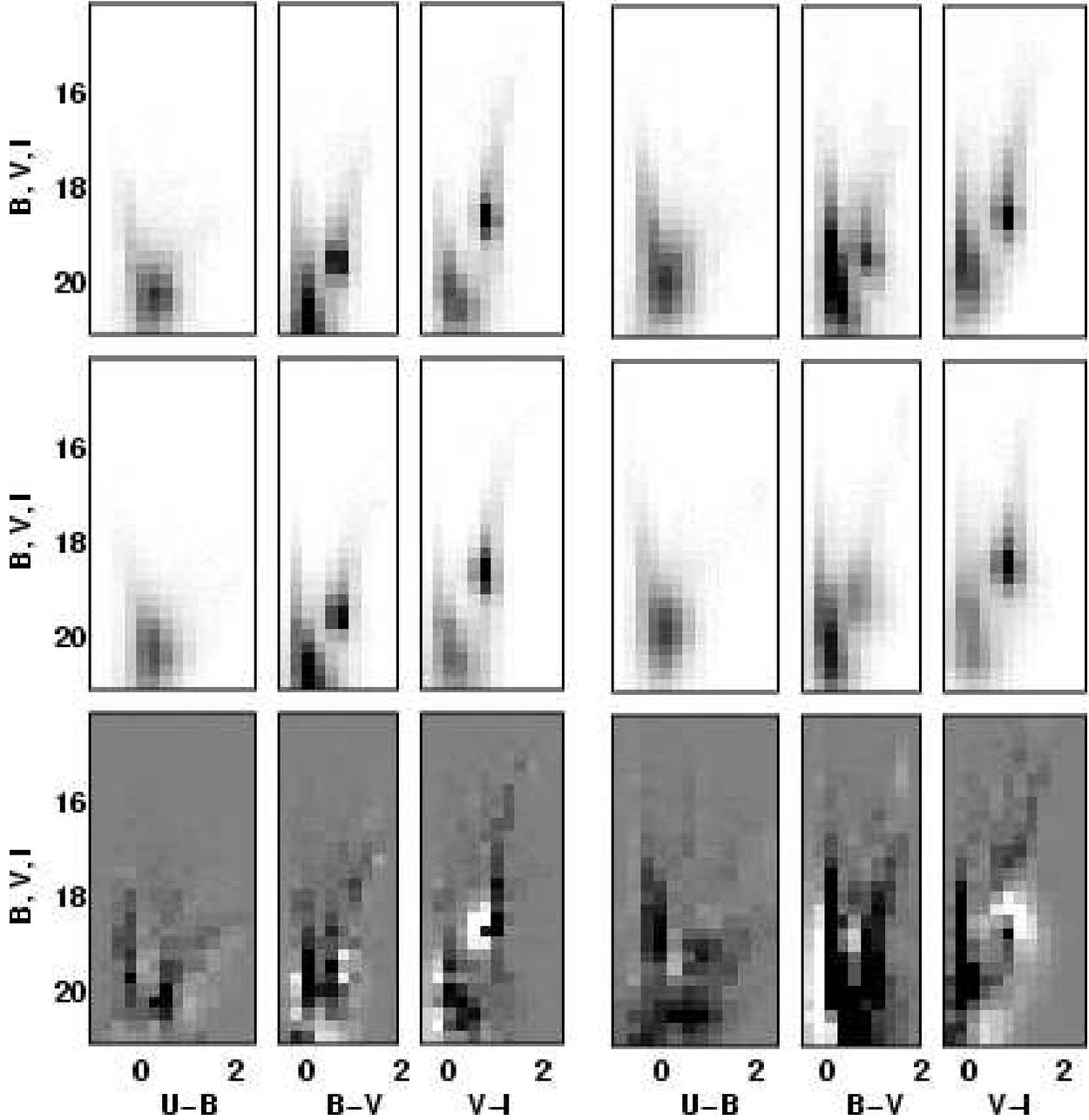}
\caption{CMD triptychs from two regions in the SMC.  At left, we show
CMDs for the region labeled MK in Figure~\ref{fig:grid} (the same
region as in Figure~\ref{fig:sfhMK}), which was well-fit by the
StarFISH algorithm, with a reduced $\chi^2$ value of 3.  At right,
we show CMDs for region KK, which was initially poorly-fit by
StarFISH, with a reduced $\chi^2$ value of 11.5.  In each case,
the top triptych shows the data photometry, the middle row shows the
best-fit model photometry, and the bottom row shows the data-model
difference (with black indicating an excess of data stars, and white
indicating an excess of model stars).  The $B-V$ difference CMD for
region KK illustrates the need for a $B-V$ offset, as discussed in
Section~\ref{sec:bad}. After applying the $B-V$ offset, the reduced
$\chi^2$ value for region KK drops to 2.7.  \label{fig:triptych} }
\end{figure}

For approximately 50 regions, the best-fit remained poor, even after
trying alternate ASTs (see Figure~\ref{fig:triptych}).  By comparing
the observed CMDs in these regions to the best-fit model triptychs,
it is apparent that the reason for the poor fit is a systematic color
offset of 0.1--0.2~mag in $B-V$ (see Section~\ref{sec:ring} for
details).  It is difficult to explain this $B-V$ color excess
astrophysically.  If the color shift is due to extinction or a
metal-rich stellar population, the $U-B$ and $V-I$ colors would have
similar color excesses, but they do not.  In fact, when we run
StarFISH with the $B-V$ CMD excluded, the solutions have much smaller
reduced $\chi^2$ values ($\sim3$--4, rather than $\sim10$).  Because
the $U-B$ and $V-I$ CMDs are well-fit by the models, we are reluctant
to apply a $B$ or $V$ offset to correct the $B-V$ CMD (because one of
the other CMDs would also be affected).  For now, we empirically apply
a $B-V$ correction to force the main sequence in these regions to lie
coincident with the main sequence of the well-fit regions.
Running StarFISH again on the problem regions after applying the
empirical $B-V$ correction, the reduced $\chi_\nu^2$ values are
dramatically lower, and are similar to those obtained in the rest of
the map. Unless otherwise noted, we will hereafter adopt the SFH
results for these regions after having applied the empirical $B-V$
correction. This correction is needed in $<$5\% of the regions, does
not affect the qualitative nature of the derived SFH in these
regions (see Section~\ref{sec:ring}), and does not affect our global
conclusions.

\subsubsection{Line-of-Sight Depth}\label{sec:losd}

Several authors have investigated the line-of-sight structure of the
SMC, generally finding a measurable extension along the line-of-sight,
although the depth measurements range from 10\% to almost 30\% of the
SMC's distance from the Milky Way.  \cite{wel87} used 91 Cepheid
variables throughout the SMC to determine a depth of $\sim7$~kpc.
\cite{mar89} examined distance moduli to young SMC stars, determining
that their depth was $<10$~kpc.  \cite{hh89} and \cite{gh91} used the
luminosity dispersion of the red clump to infer depths as large as
17~kpc in the outer regions of the SMC (although perhaps half as large
as this in many regions).  More recently, \cite{gro00}
cross-referenced hundreds of cepheids in the OGLE, DENIS, and 2MASS
surveys, finding a depth of 14~kpc; while \cite{cro01} used SMC star
clusters to determine a depth of 6--12~kpc.

To test whether we need to account for line-of-sight depth in our
analysis, we assume a characteristic depth of 12~kpc, corresponding to
$\pm0.2$~mag in distance modulus.  We constructed artificial stellar
populations with this intrinsic luminosity spread, and performed the
StarFISH analysis {\it without} accounting for the spread in distance.  
The $\chi^2$ value of the solution is slightly inflated compared to an
identical zero-depth population, but the SFH solutions were the same,
within the errors.  In addition, we see no empirical evidence in the
zero-depth model fits to the real data that a significant
line-of-sight depth is required, so we omit it in the present
analysis.

\subsection{Results: A Map of the SMC's Star Formation
History}\label{sec:map}

In Table~\ref{tab:smcsfh}, we present our best SFH solutions for 351
regions in the SMC.  The SFH amplitudes output by StarFISH are equal
to the number of stars which were formed in each age/metallicity bin.
In the Table (and in all subsequent discussion), the SFH amplitudes
have been converted to star-formation rates by simply multiplying by
the IMF-dependent mean stellar mass, and dividing by the age interval
covered by the bin.  This is straightforward for all bins except the
oldest, for which we have only a lower age limit.  We adopt an upper
age limit of 13.7 Gyr in computing the star-formation rates of the
oldest bin.

The information presented in Table~\ref{tab:smcsfh} is condensed into
a map of the SMC's star-formation history in Figure~\ref{fig:sfhmap}.
Each panel in the Figure represents a ``snapshot'' of the star-
formation activity at a particular epoch.  In each panel, the 351
subregions are represented as ``pixels'' of variable size whose
brightness is proportional to the local star-formation rate (SFR).  
The SFH map is available as an animation at [URL to be specified].

\section{Discussion}\label{sec:discuss}

\subsection{The Global Star Formation History}\label{sec:sfh}

In Figure~\ref{fig:totsfh}, we show the global SFH of the SMC,
derived by summing together the star-formation rates over all 351
subregions and over all three metallicities.  The SFH revealed by
Figures~\ref{fig:sfhmap} and \ref{fig:totsfh} contains several
interesting features:
(1) there was a significant epoch of star formation in our oldest
age bin, covering all ages older than 8.4~Gyr,
(2) there was a long quiescent epoch between 3 and 8.4~Gyr ago,
during which the SMC apparently formed relatively few stars,
(3) the quiescent epoch was followed by more-or-less continuous star
formation starting about 3~Gyr ago, and extending to the present,
(4) superimposed on the recent continuous star formation, there are
at least three peaks in the SFR, at 2--3~Gyr, 400~Myr and 60~Myr ago,
and
(5) there is a ring-like morphology in the intermediate-age frames
(2.5--1.0~Gyr) that may suggest an inward propagation of star
formation or the remnant of a gas-rich merger event.

\subsubsection{The First Stars}\label{sec:oldstars}

Although it is difficult to discern any details about the SFH at the
earliest times, the determination that a large fraction of all stars
in the SMC correspond to a population from the earliest times is
critical. We find that the SMC formed about 50\% of its total stellar
population prior to 8.4~Gyr ago, or alternatively at $z > 1.2$
for WMAP cosmology \citep{spe03}.  Like all Local Group systems
\citep{mat98}, the SMC contains a significant old population.

It is useful to compare our results for the early history of the SMC
to previous work in this area.  \cite{gh92} found that the bulk of the
SMC's stars are aged $\sim10$~Gyr; while they didn't offer a
quantitative fraction, from their discussion it can be inferred that
the number is substantially larger than 50\%.  However, this
difference may be attributed to the fact that \citeauthor{gh92}
studied the outer portions of the SMC where the younger stellar
populations are probably much less common.

More recently, \cite{dol01} used a deep {\it HST WFPC2}
field near NGC~121 to reconstruct the old SFH of the SMC.  The deep
{\it HST} photometry includes stars well below the ancient main
sequence turn-off, so it is a superior data set for reconstructing
the early history of the SMC in this regard.  \citeauthor{dol01} find
a SFH which peaks between 5 and 8~Gyr ago, in contrast to
what we find.  However, there are two factors which make direct
comparisons of the two solutions difficult.  First, our analysis
regions do not overlap; they used a field near NGC 121, which falls
within the region we masked out to avoid contamination from the
foreground galactic cluster 47~Tucanae.  Second, the \citeauthor{dol01}
field is well outside the main portion of the SMC, so it would seem
dangerous to infer a general SFH for the entire SMC based on this
small field at its periphery.  Still, one could argue against both
of these explanations by pointing out that for the old SFH at least,
the stellar populations should be well-mixed, so even a tiny sample
should be representative of the whole.  Part of the problem may be the 
coarser age resolution employed in our analysis (as mandated by our 
ground-based data).  More deep-field SFH analyses from different regions 
in the SMC should be performed to investigate the issue.

\subsubsection{The Quiescent Epoch}\label{sec:quiescent}

The second and third panels of Figure~\ref{fig:sfhmap} are globally
dark, indicating an epoch lasting several Gyr during which the SMC
formed relatively few stars.  To ensure that the lack of detected
star formation in these panels is not an artifact of our method, we
added supplemetary synthetic stellar populations to several selected
SMC regions.  The supplemental populations have the correct number
and age distribution (3--8.4~Gyr) to fill in the apparent age gap in
the observed SFH.  We found that the StarFISH algorithm successfully
recovered the ``observed + synthetic'' star formation history,
indicating that the method is not inherently insensitive to stellar
populations in this age range.

The quiescent epoch we infer from the SMC's field stars is
intriguingly coincident with the well-known ``age gap'' among star
clusters in the Large Magellanic Cloud \citep{vdb91, dc91, gir95, 
wes97}.  Although the cluster population in the SMC has been 
conventionally regarded as having a continuous age distribution
\citep{dh98, msf98}, analysis of  deep {\it Hubble Space Telescope}
photometry of the SMC's seven brightest old (age $>1$~Gyr) clusters
\citep{rich00} has shown that most of the SMC's old clusters were 
formed in two sharply distinct episodes: one that occurred 
$8\pm2$~Gyr ago and one that occurred $2\pm0.5$~Gyr ago.
Furthermore, \citeauthor{rich00} discuss three SMC clusters that are 
too faint to be included in their $HST$ study.  One (Lindsay~1) has a 
ground-based age of 9~Gyr, coincident with the older burst.  The 
other two have ground-based ages between the 2 and 8~Gyr bursts 
(Lindsay~11 has an age of 3.7~Gyr and Lindsay~113 has an age of 
6~Gyr).  The age of Lindsay~113 was confirmed by \cite{cro01}, who
used ground-based photometry to derive an age of $5.3\pm1.3$~Gyr.  
Therefore, our present understanding is that among the SMC's ten 
populous old clusters, eight were formed in one of two epsiodes, and 
two were formed at some time between the episodes.  While the number 
of SMC clusters is small, their age distribution is not uniform and
is qualitatively consistent with our SFH derived from the SMC's field 
population. In a subsequent paper, we will present an analysis of the 
ages of 204 SMC stellar clusters that independently confirm the 
quiescent epoch seen here among the stars, as well as the bursts 
discussed next.

\subsubsection{``Bursts'' of Star Formation}\label{sec:bursts}

From Figure~\ref{fig:totsfh}, the SFH since 3~Gyr ago can be
characterized as having an underlying constant SFR of
$\sim0.1\ M_\odot\ yr^{-1}$ with superimposed episodes of enhanced
star formation at 2--3~Gyr, 400~Myr, and 60~Myr.  The mean SFR in
these age bins is a factor of 2--3 times higher than in the
surrounding bins.

We hesitate to conclude that the SMC has had exactly three bursts in
its history, because rapid star-formation events can happen on 
timescales that are orders of magnitude shorter than synthetic CMD 
methods are able to resolve, especially at ages $\simgreat1$~Gyr.
The actual star formation rate as a function of time could be varying
wildly within any of our age bins, and we would not know it. We only
measure the mean star formation rate over the width of the bin.  It
is certainly possible that these enhanced star-formation events were
dominated by a multiple, distinct, short-duration burst, which were
each much stronger than the 2--3$\times$ enhancement reflected in the
mean star formation rate.  Similarly, we do not intend to imply that
we have actually observed a more-or-less constant inter-burst SFR in 
the SMC over the past 3~Gyr; the unknowable possibility of short-term
SFR variations prevents such a conclusion.  We simply characterize
the observed SFH as having a baseline constant SFR in order to
highlight the three superimposed episodes of heightened star
formation.

Figure~\ref{fig:totsfh} indicates 5 times at which the SMC had a
perigalactic encounter with the Milky Way, and 14 perigalactic
encounters with the LMC, over the past 12 Gyr \citep{ljk95}.
In addition, the SMC is believed to be currently very near
perigalacticon with respect to the Milky Way.  The most recent
perigalactic encounters ($\sim500$~Myr ago for the LMC, $\sim2.5$~Gyr
ago for the Milky Way) fall in the age bins in which we have observed
significantly enhanced star formation rates, raising the possibility
that we have recorded the effects of interaction-induced star
formation in the SMC.  For the older encounters, we lack the age 
resolution to discern any response in the SMC's SFH to these 
short-lived events.  These issues are explored in greater 
detail in a companion paper \citep{zh04}.

\subsubsection{A Ring of Star Formation?}\label{sec:ring}

\begin{figure}
\plotone{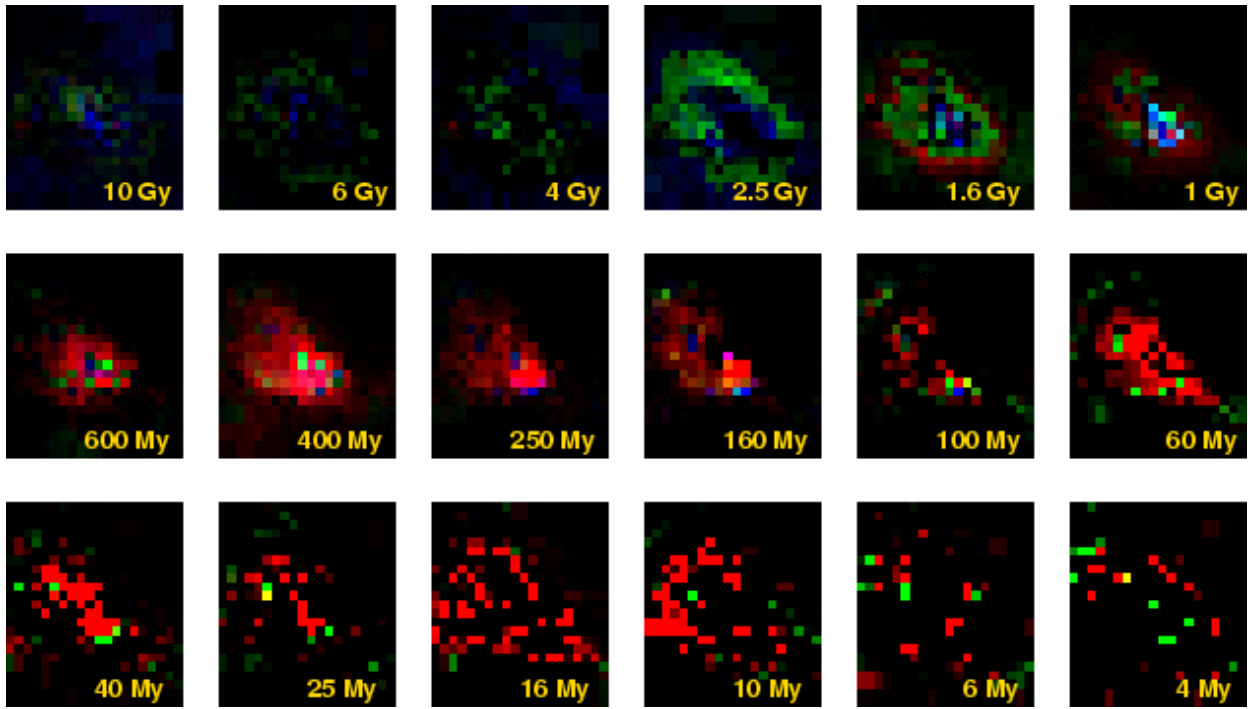}
\caption{The star-formation history of the SMC.  Each panel shows the
star formation activity for a particular age bin; the characteristic
age of the bin is labeled.  Each pixel represents one of our 351 SMC
subregions (see Figure~\ref{fig:grid}), with a pixel value that is
proportional to the subregion's star formation rate (in the electronic
edition, the pixels are also color-coded to reflect the mean
metallicity of the stars.  Red corresponds to Z=0.008, green
corresponds to Z=0.004, and blue corresponds to Z=0.001).
\label{fig:sfhmap} }
\end{figure}

\begin{figure}
\plotone{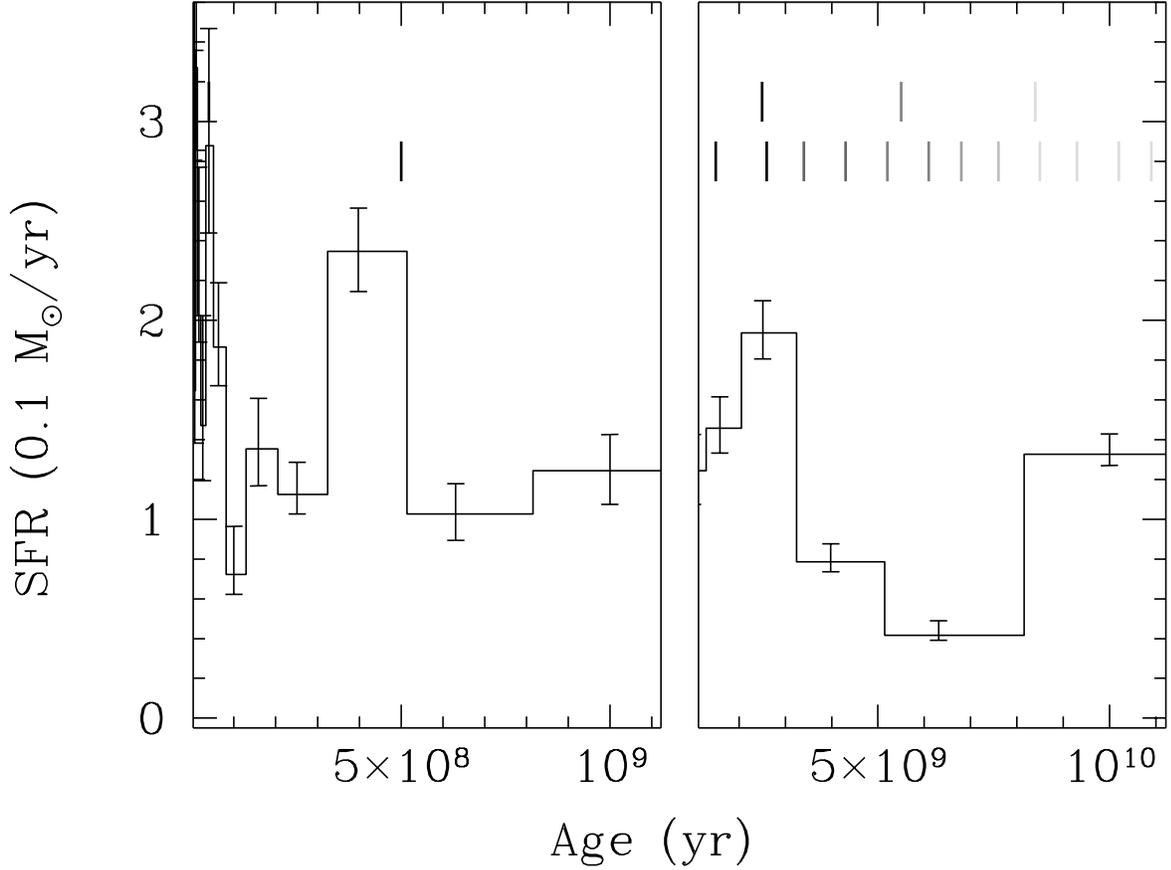}
\caption{The global star formation history of the SMC.
We sum the star formation rates across metallicity and the 351
individual subregions.  We use a split-panel view with
different scalings of the time axis because the age bins become
logarithmically narrower for younger ages.  The displayed error bars
represent the 1-$\sigma$ confidence interval for each age bin and
include covariance between age bins.  The times at which the SMC had a
perigalactic passage are indicated by the rows of short vertical lines
\citep{ljk95}.  The top row indicates encounters with the LMC, the
bottom row indicates encounters with the Milky Way.  These lines fade out
at older ages as a representation of the uncertainty of the  
encounter times with increasing age. \label{fig:totsfh} }
\end{figure}

There is unexpected large-scale spatial structure in the SFH map
(Figure~\ref{fig:sfhmap}) at intermediate ages (2.5 to 1~Gyr).
Naively, one might expect stellar populations of this age to be
well-mixed and that their distribution would follow the overall
stellar density (see Figure~\ref{fig:grid}).  The ring is most
prominent in the 2.5~Gyr frame, where the dense central regions are
almost totally quiescent, and the encircling ring is highly active,
especially to the northeast.  In the 1.6~Gyr frame, there is some
low-metallicity star formation activity in the central regions, and
the active ring regions have become more metal rich.  Starting with
the 1~Gyr frame, the distribution of SFRs is finally centrally peaked
as one would expect, but the activity in the central regions is still
of lower metallicity than in the encircling ring.

Before we attempt to interpret this structure, we must determine
whether it is an artifact in the data or of the analysis.  There are
three reasons to suspect that the ring may not be real.  First, the
shape of the ring approximately follows the stellar surface density
contours in the SMC.  If our method is subtly sensitive to errors in
the modeling of the stellar crowding, we might expect to see
recovered age discrepancies proportional to the projected stellar
density, which could result in a ring-like artifact in the SFH map.
Second, and perhaps related to the first point, the regions that form
the ``hole'' interior to the ring feature are the same regions which
required a modest $B-V$ color offset for us to obtain an acceptable
SFH model.  Third, the synthetic CMDs corresponding to the age range
where the ring is most prominent have their main-sequence turn offs
at the faint end of the CMD, where the completeness rate and
photometric errors change rapidly as a function of magnitude and are
most difficult to model.

\begin{figure}
\plotone{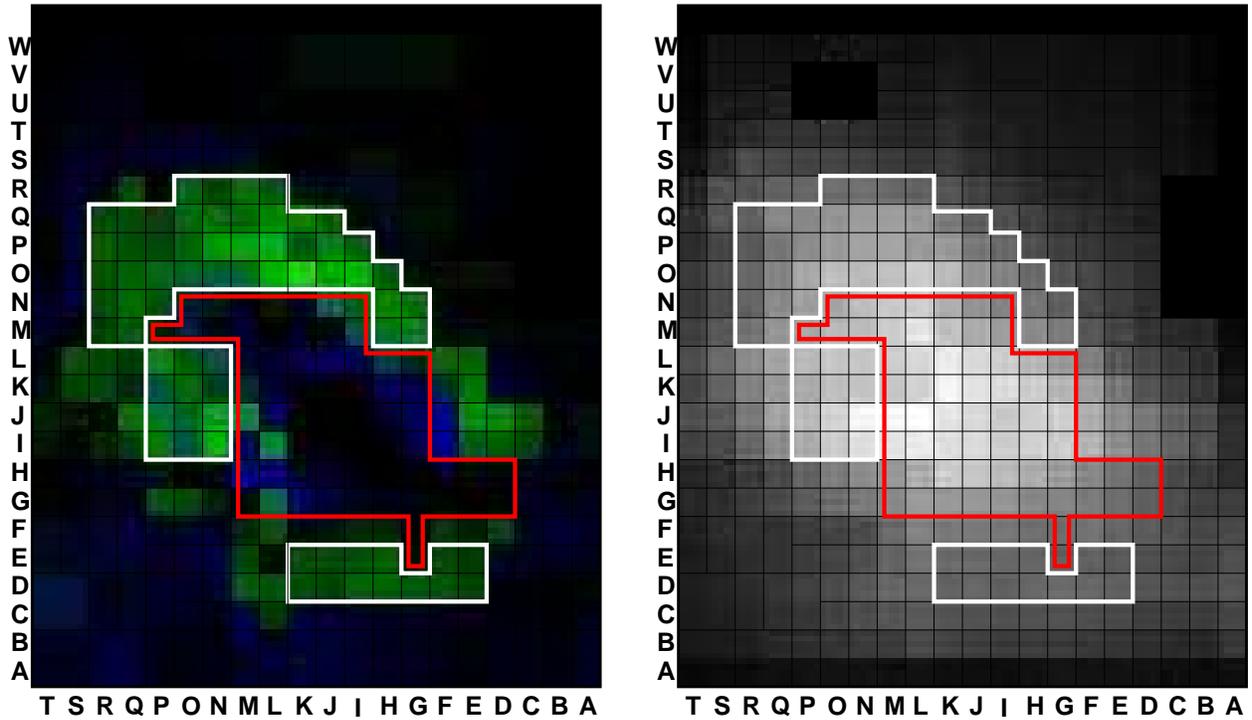}
\caption{The left panel shows the 2.5~Gyr frame of our SFH map
(Figure~\ref{fig:sfhmap}), the right panel shows the star counts
in each subregion (Figure~\ref{fig:grid}).  We highlight the 68
regions that we combine into an ``on'' population, based on their
star formation rate at 2.5~Gyr (solid white outline).  We also
highlight the 68 regions that we combine into an ``off'' population
(dashed outline; or red outline in the electronic edition).  These
two composite populations are compared to investigate the nature of
the ring-like structure visible in the 2.5~Gyr map frame.
\label{fig:ring} }
\end{figure}

\begin{figure}
\plotone{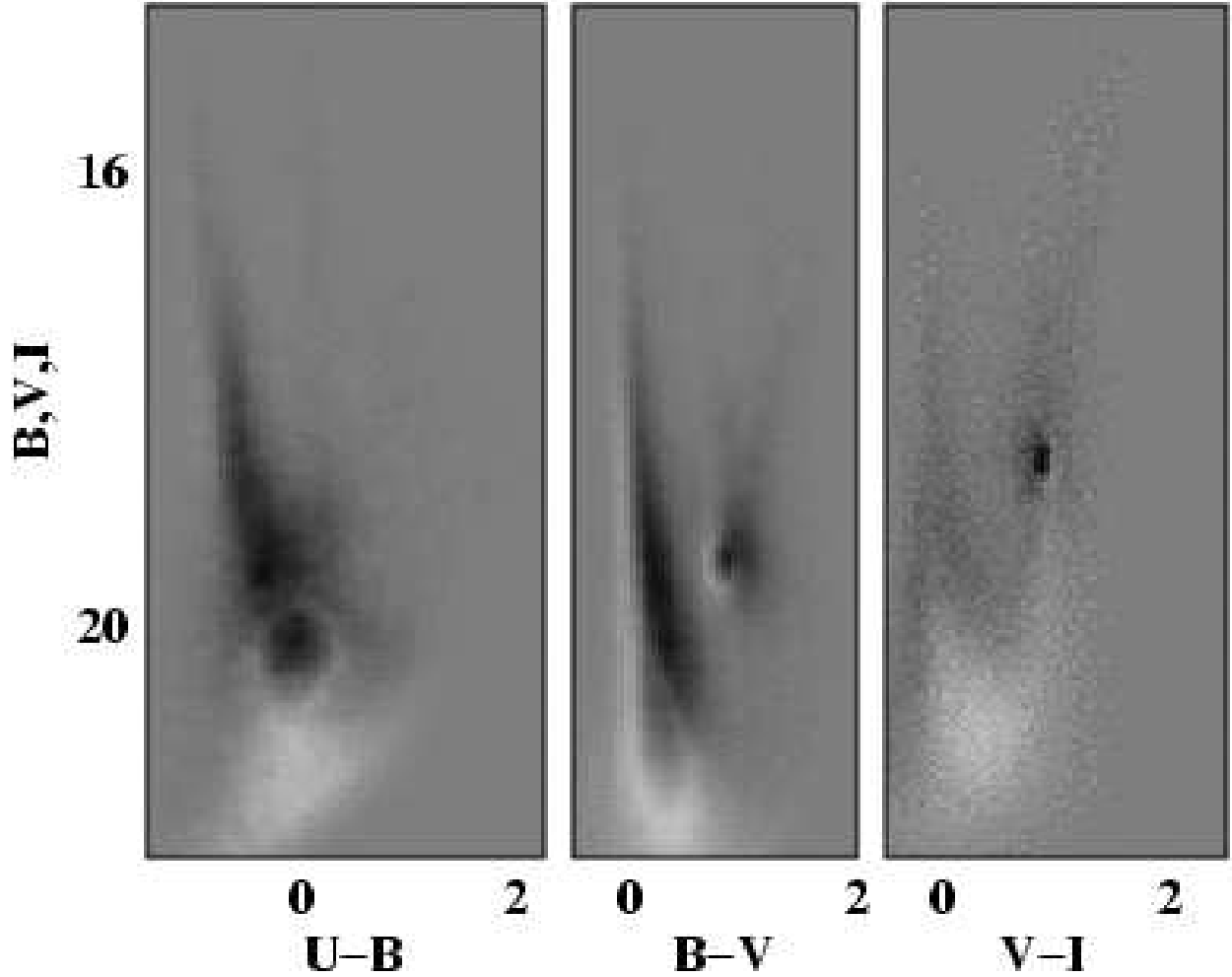}
\caption{CMD triptych showing the difference between the ``on'' and
``off'' meta-regions.  White pixels indicate an excess
of stars among the ``on'' population; black pixels indicate an excess
of stars among the ``off'' population. The ``on'' population has an
excess of faint main-sequence stars, compared to the ``off''
population, and there is a systematic offset in the $B-V$ color of
the two populations. \label{fig:ring-cmds} }
\end{figure}

To understand which features in the CMDs are driving the dramatic
contrast between the active and quiescent regions which form the
ring at 2.5~Gyr, we use the 2.5~Gyr SFH frame of
Figure~\ref{fig:sfhmap} to isolate and compare two stellar
populations: an ``on'' population drawn from the active regions
within the ring, and an ``off'' population drawn from the quiescent
central regions (see Figure~\ref{fig:ring}).  CMD triptychs for these
composite regions, as well as the difference between them, are shown
in Figure~\ref{fig:ring-cmds}.  These CMDs show the original
photometry, {\it without} the applied $B-V$ offset that was used in
determining the SFHs (see Section~\ref{sec:bad}).

The most prominent feature of the difference CMDs is an excess of
faint main sequence stars in the ``on'' regions relative to the
``off'' regions. Because the central regions are also generally
the most crowded, this difference can be regarded as the expected
result of a brighter faint limit in the more crowded regions.
Although this difference should not affect the best-fit SFHs if
our ASTs are correct, it is suspicious that that this difference
lies precisely where the main sequence turn-off stars of a 2--3~Gyr
population would be found.  Overestimating the faint-end completeness
rate in the central regions would result in a suppressed formation
rate at 2--3~Gyr, similar to that observed in the central regions.

If the deficit of faint main sequence stars is the explanation for
the suppresed star formation in ``off'' regions, we must understand
why the ASTs failed to produce a viable model of the completeness
rate in these regions.  The problem cannot simply be due to higher
stellar surface densities of the ``off'' regions because there are
``on'' regions which have similarly high stellar surface densities
(see Figure~\ref{fig:ring}).  If there is an AST failure, it is more
likely caused by our assumption that crowding effects dominate the
photometric errors; in other words, there may be non-crowding
parameters affecting the photometry of the ``off'' regions.  This
will skew our results if the SFH solution uses a synthetic CMD
library based on ASTs from an image which is not an ``off'' region.
However, even this cannot be the right answer, because we are in fact
using ASTs derived from ``off'' regions to construct their synthetic
CMD libraries.  In particular, we use ASTs from regions JJ and KK, so
at the very least, these two regions must have photometric error
characteristics that are appropriately described by the adopted ASTs.
Yet regions JJ and KK are unequivocally among the ``off'' regions.
We find no reason to believe that these AST results are in error.

\begin{figure}
\plotone{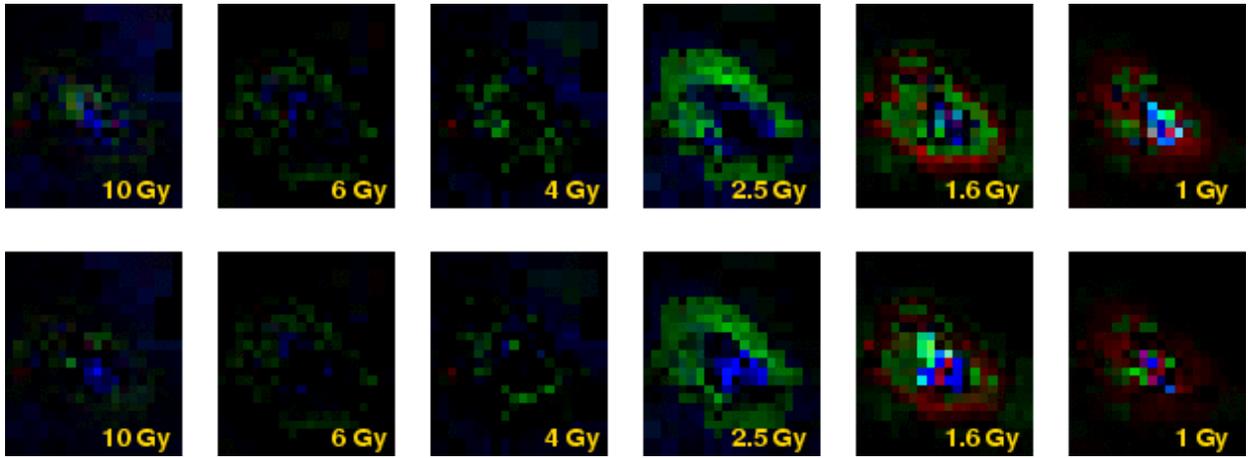}
\caption{Comparing the oldest six bins of the SFH map, with and
without the $B-V$ color offset.  Top row: the original solution,
in which we applied a $B-V$ offset to some of the subregions, in
order to improve the $\chi^2$ values of our SFH solutions (see
Section~\ref{sec:bad}).  Bottom row: An alternate solution, in
which we did not apply color offsets to any subregions.
\label{fig:comparemap} }
\end{figure}

\begin{figure}
\plotone{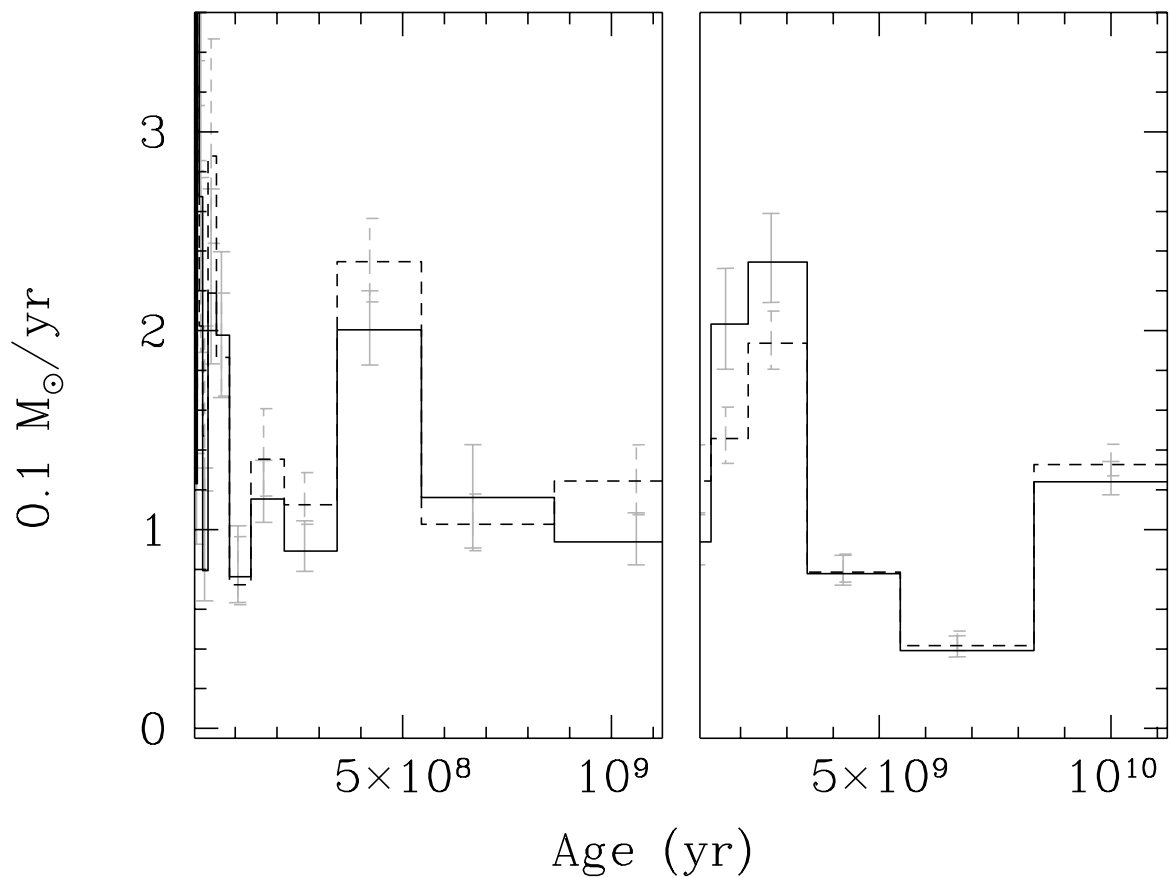}
\caption{Comparing the global SFH solution, with and without the
$B-V$ color offset.  The original SFH solution from
Figure~\ref{fig:totsfh} is shown as the solid histogram.  This
solution included a $B-V$ color offset applied to the photometry
of some subregions to improve the $\chi^2$ of their SFH solutions.
The dotted histogram shows an alternate SFH solution, for which no
$B-V$ color offsets were applied.  There is no significant
difference between the two solutions.
\label{fig:comparesfh} }
\end{figure}

Another notable feature visible in the difference CMDs of
Figure~\ref{fig:ring-cmds} is the displacement of the main sequence
in the $B-V$ CMD, in the sense that the main sequence in the ``off''
regions is systematically redder than that of the ``on'' regions.
The most plausible explanation seems to be a systematic photometric
zeropoint offset in the photometry of these central regions, which
motivated us to apply a $B-V$ correction to improve their best-fit
SFHs (see Section~\ref{sec:bad}).  However, since the cause of this
offset is unknown, we must suspect the SFH solutions for these
regions.  We investigate the effect of the $B-V$ color correction on
the SFHs of the ``off'' regions by comparing the SFH solutions
including the $B-V$ correction to a second set of SFH solutions in
which the $B-V$ correction was not applied (see
Figure~\ref{fig:comparemap}).  The ``hole'' is less prominent when no
color offset is applied, but it is still present, and there is still
a strong radial trend in metallicity.  In
Figure~\ref{fig:comparesfh}, we compare the global SFH solutions for
the cases with and without the applied $B-V$ offset.  This Figure
shows that the application of the $B-V$ offset to the central
subregions has very little effect on our overall SFH solution for
the SMC.  Because the $\chi^2$ values are substantially improved when
the $B-V$ offset is applied, we retain our original SFH solutions,
which include the offset.  The nature of the $B-V$ offset remains a
mystery and so the SFH of the central regions must be viewed with
caution.  Further understanding of these apparently anomalous main
sequence populations may require deeper photometry with large
ground-based telescopes or with the $HST$.

If the observed ring feature is real, it suggests the possibility of
a global inward propagation of star formation in the SMC on a
timescale of a few Gyr.  Alternatively, the ring may be composed of
stars that formed as the result of a gas-rich merger 2--3~Gyr ago.
The latter is consistent with an infall scenario that reproduces the
SMC's chemical enrichment history (\cite{zh04}) and the different
distribution of young and old stars in the SMC \cite{zar00}. In this
scenario, the kinematics of the stellar population formed by the
merger would follow the kinematics of the infalling gas; the gas
kinematics might have sufficient angular momentum to produce stellar
populations in a persistent annular distribution. Kinematic
measurements of different populations in the SMC might
test this scenario.

\subsection{The Chemical Enrichment History}\label{sec:ceh}

Previous analyses of the chemical enrichment history (CEH) of the
SMC, based primarily on measurements of $\sim10$ stellar clusters
in the SMC \citep{dop91}, have noted very little change in the
metallicity of clusters between 10 and 4~Gyr old. To explain this
observation, investigators have invoked either significant infall of
unenriched gas or a ``leaky box'' model, in which supernova-driven
winds preferentially remove heavy elements. However, \cite{dh98}
re-examined the cluster data, and concluded that if two ``anomalous''
clusters are ignored, the age-metallicity relation increases
gradually and monotonically, consistent with a simple closed-box
enrichment model.  \cite{pt99} again examined the cluster data and
concluded, in agreement with \cite{dop91}, that the metallicity in
the SMC remained low until $\sim4$~Gyr ago.  \citeauthor{pt99}
present a simple chemical enrichment model consistent with the
cluster data, in which the long period of stagnant enrichment
between 4 and 10~Gyr is explained by a lull in the star formation
rate over the same period.  We note that we have observed a very
similar lull in the star formation rate among the SMC's field
populations (see Figure~\ref{fig:totsfh} and
Section~\ref{sec:quiescent}).

In addition to the work on star clusters, field variable stars
have also provided some constraint on the CEH of the SMC.
\cite{smi92} inferred from the period-amplitude relation that the
metallicity of RR~Lyrae stars in the SMC is similar to that of the
field giants of \cite{sun86}; [Fe/H]=-1.6.  \cite{but82} measured the
metallicities of three field RR Lyrae stars directly from spectra,
finding $<[Fe/H]>=-1.8$ for these old stars.  In addition, \cite{har81}
measured photometric abundances of 45 Cepheid variables in the SMC,
finding $<[Fe/H]>=-0.5$ for these intermediate-age ($\sim10^8$~yr)
populations.

Our recovery of the SMC's SFH provides an independent determination
of its CEH. Before beginning the SFH analysis, we suspected that we
might have to impose {\it a priori} constraints on the CEH, since
stellar photometry allows only crude measurements of metallicity.
Instead, we found that StarFISH was able to converge upon a
reasonable CEH without {\it a priori} constraints.

\begin{figure}
\plotone{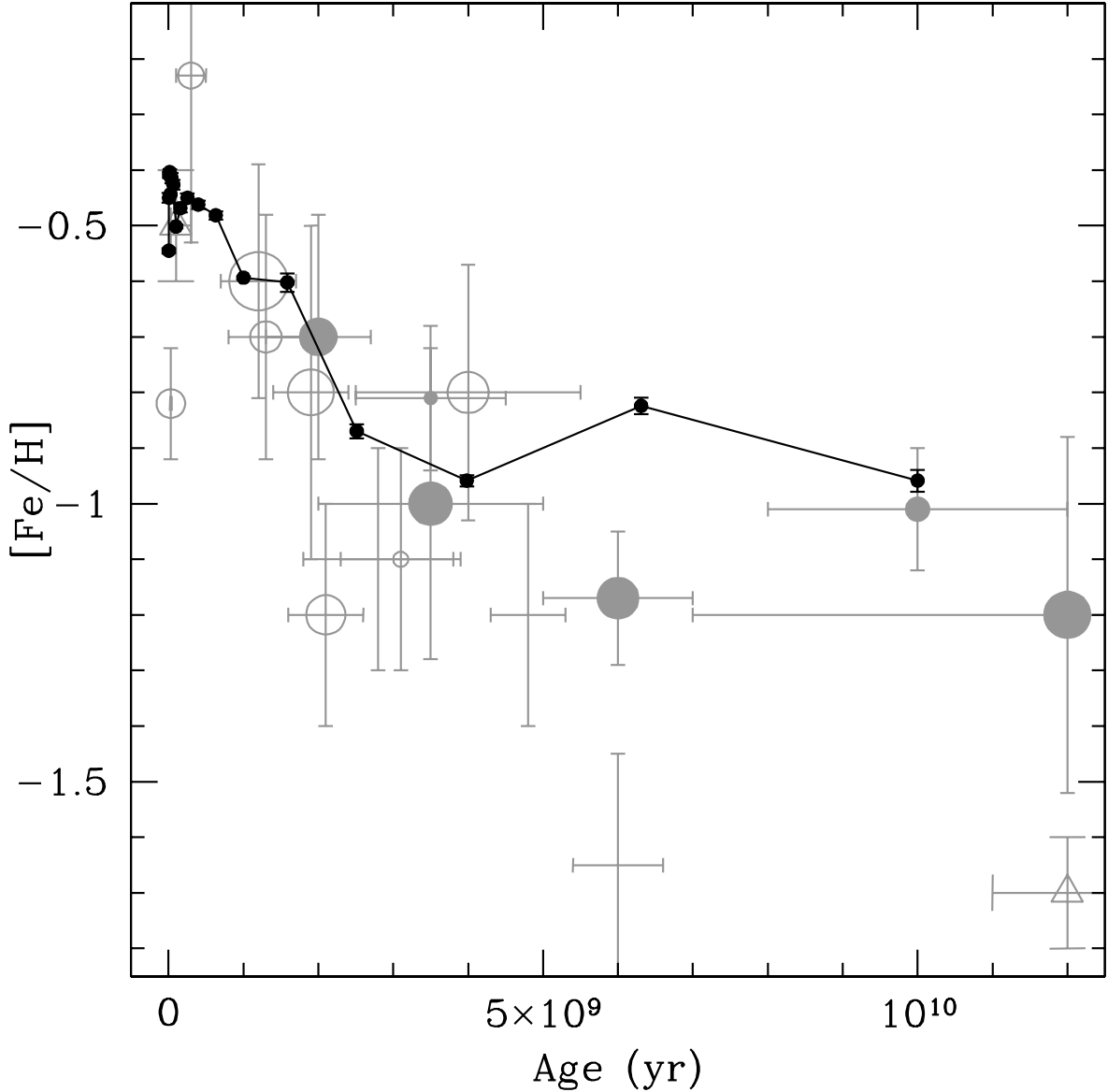}
\caption{The age-metallicity relation derived from our SFH analysis
is shown as the small points connected with straight-line segments.
The grey points represent existing metallicity and age measurements in
the literature.  Circular points represent star clusters, and the size
of each circle is proportional to the cluster's estimated mass (some
circles are too small to be seen in the Figure).  The six solid
circles represent the clusters used by \cite{dop91}, \cite{dh98} and
\cite{pt99}, and the open circles are from \cite{dfp98} and
\cite{pia01}.  The triangular point at 11~Gyr represents the field RR
Lyrae stars as measured by \cite{smi92} and \cite{but82}, and the
triangular point at 0.1~Gyr represents the field Cepheid stars, as
measured by \cite{har81}. \label{fig:agez} }
\end{figure}

To quantify the CEH, we determine the mean metallicity of stars
formed at a given age, and plot this as a function of age in
Figure~\ref{fig:agez}.  We convert from $Z$ metallicities to
$[Fe/H]$ values by adopting $Z_\odot=0.02$, and assuming that
$[Fe/H] = log(\frac{Z}{Z_\odot})$.  The error bars on our
age-metallicity relation (AMR) represent the standard deviation of
the individual subregions' metallicity values about the global mean
metallicity in each age bin, and do not necessarily reflect the
precision with which we have determined the mean metallicity.

We find that our AMR agrees quite well with the star cluster and
field variable data, even though our photometrically-derived
metallicities are necessarily crude.  We find that the field
metallicity in the SMC remained rather low ($[Fe/H]\sim-1.0$) until
2--3~Gyr ago, at which point it began a steady increase to its
present-day value of $[Fe/H]\sim-0.5$.  There is a marginal indication
that the mean metallicity actually {\it decreased} between 7 and 4~Gyr
ago.  Our derived CEH is consistent with the chemical enrichment model
of \cite{pt99}, lending further support to the presence of a
quiescent epoch in the SMC's early history.  The implications of the
CEH in the context of the SMC's interaction history is discussed in
detail in a subsequent paper \citep{zh04}.

\section{Summary}\label{sec:summary}

From the reconstruction of $UBVI$ color-magnitude diagrams using
the StarFISH \cite{hz01} analysis package and comparison to
the MCPS $UBVI$ photometry of stars in the SMC \cite{zar02}, we
determine the global star formation history as resolved on a grid of
351 independent subregions. Critical components of this analysis
include differential reddening that accounts for both spatial and
population dependencies, and extensive artificial star tests to
determine the photometric error model and completeness correction.

We find that the recovered SFH of the SMC can be divided into three
epochs:

\noindent
1) An early epoch ($t>8.4$~Gyr ago) where a significant fraction
($\sim50$\%) of all stars in the SMC were formed. Because we have
poor temporal resolution at these early times, we can only constrain
the total number of stars older than 8.4~Gyr, and not a more detailed
distribution of their ages.

\noindent
2)  An intermediate epoch ($3 < t < 8.4$~Gyr) where the SMC
experienced a long quiescent period during which it formed relatively
few stars. This quiescent time parallels the lack of known clusters of
these ages both in the LMC and SMC.

\noindent
3) An active recent time ($t < 3$~Gyr) where there has been
continuous star formation punctuated by ``bursts'' at 2.5~Gyr,
400~Myr and 60~Myr.  The older two events are coincident with past
perigalactic passages by the SMC with the Milky Way.  The strongest
burst, that at 2.5 Gyr, appears to have an annular structure and
an inward propagation spanning $\sim1$~Gyr.  However, we remain
skeptical about the reality of this structure, and
deeper photometry of the SMC's crowded central regions is
required to investigate further.

For our derived chemical enrichment history we find:

\noindent
1) that the mean chemical abundance of stars formed in the SMC
remains low ($[Fe/H] \sim -1$) until about 3 Gyr ago and then
rises monotonically to the present gas-phase abundance value of
$[Fe/H] \sim -0.4$].

\noindent
2) that our observed AMR is consistent with existing age and
metallicity measurements of populous star clusters in the SMC.

\noindent
3) that our relation is inconsistent with a simple
closed-box enrichment model, unless the quiescent epoch we have
observed was nearly devoid of significant star formation.

Because of the decreasing temporal resolution with lookback time,
it is not straightforward to infer the detailed behavior of the
star formation rate from the reconstructed SFH. Instead,
models must be convolved with the binning structure imposed by
the method. In a companion paper \cite{zh04} we address whether
the SFH and chemical enrichment history is consistent with a model
where pericenter passages drive star formation. Both the SFH and
chemical enrichment history are sufficiently complex and rich in
behavior that we hold out hope that they may be able to provide
interesting constraints on the nature of star formation in
galaxies.

\vskip 1in
\noindent Acknowledgements:
DZ acknowledges financial support from National Science Foundation
CAREER grant AST-9733111 and a fellowship from the David and Lucile
Packard Foundation.

\clearpage


\clearpage

\input{jharris.tab1}
\input{jharris.tab2}

\clearpage

\end{document}

%% file: jharris.tab1.tex
\begin{deluxetable}{rrrr}
\tablecaption{The Artificial Stars Tests \label{tab:ast}}
\tablewidth{0pt}
\tablehead{ 
  \colhead{Subregion\tnm{a}} & 
  \colhead{$N_\star/\Box\arcmin$} & 
  \colhead{Subregion\tnm{a}} & 
  \colhead{$N_\star/\Box\arcmin$} 
}
\startdata
TM &  57.4 & MO & 176.6 \\ 
RM & 104.5 & MM & 186.4 \\ 
PM & 152.1 & JJ & 197.6 \\ 
PO & 153.5 & KK & 221.6 \\ 
\enddata
\tablenotetext{a}{Each two-letter code indicates the region's position
in our gridding of the SMC (see Figure~\ref{fig:grid}).}
\end{deluxetable}

%% file: jharris.tab2.tex
\begin{deluxetable}{rrrrrrrrrr}
\tabletypesize{\scriptsize}
\tablecolumns{10}
\tablewidth{0pt}
 
\tablecaption{The Star Formation History of the SMC \label{tab:smcsfh}}
\tablehead{
    \colhead{Age Range} & \multicolumn{3}{c}{Z = 0.008} & \multicolumn{3}{c}{Z = 0.004} &
        \multicolumn{3}{c}{Z = 0.001} \\ \cline{2-4} \cline{5-7} \cline{8-10} 

    \colhead{log(yr)} & 
        \colhead{$SFR$} & \colhead{$SFR_{low}$} & \colhead{$SFR_{high}$} & 
        \colhead{$SFR$} & \colhead{$SFR_{low}$} & \colhead{$SFR_{high}$} & 
        \colhead{$SFR$} & \colhead{$SFR_{low}$} & \colhead{$SFR_{high}$} 
}
 
\startdata
\multicolumn{10}{c}{Region AA  ( 0$^h$ 25$^m$,   -74\arcdeg\ 57\arcmin )} \\ 
\cline{1-10} 
9.925--10.05 &     0 &     0 &    29  &      0 &     0 &    42  &    416 &   306 &   526 \\ 
9.725--9.925 &     0 &     0 &    23  &      0 &     0 &    32  &     37 &     0 &   120 \\ 
9.525--9.725 &     0 &     0 &    23  &      0 &     0 &    42  &      0 &     0 &   100 \\ 
9.325--9.525 &     0 &     0 &    29  &      0 &     0 &    61  &    331 &   246 &   416 \\ 
9.125--9.325 &     0 &     0 &    48  &    106 &    32 &   180  &      0 &     0 &    61 \\ 
8.925--9.125 &     0 &     0 &    43  &      0 &     0 &    56  &      0 &     0 &    48 \\ 
8.725--8.925 &     0 &     0 &    50  &      0 &     0 &    52  &      0 &     0 &    53 \\ 
8.525--8.725 &    55 &     0 &   109  &      0 &     0 &    50  &      0 &     0 &    53 \\ 
8.325--8.525 &     0 &     0 &    57  &      0 &     0 &    62  &      0 &     0 &    64 \\ 
8.125--8.325 &     0 &     0 &    80  &      0 &     0 &    80  &      0 &     0 &    86 \\ 
7.925--8.125 &     0 &     0 &   110  &      0 &     0 &   110  &      0 &     0 &   120 \\ 
7.725--7.925 &     0 &     0 &   180  &      0 &     0 &   180  & \nodata & \nodata & \nodata \\ 
7.525--7.725 &     0 &     0 &   230  &      0 &     0 &   240  & \nodata & \nodata & \nodata \\ 
7.325--7.525 &     0 &     0 &   360  &      0 &     0 &   370  & \nodata & \nodata & \nodata \\ 
7.125--7.325 &     0 &     0 &   560  &    145 &     5 &   725  & \nodata & \nodata & \nodata \\ 
6.925--7.125 &     0 &     0 &   900  &      0 &     0 &   890  & \nodata & \nodata & \nodata \\ 
6.725--6.925 &     0 &     0 &  1400  &      0 &     0 &  1500  & \nodata & \nodata & \nodata \\ 
6.600--6.725 &     0 &     0 &  2600  &      0 &     0 &  2900  & \nodata & \nodata & \nodata \\ 
\enddata
\tablecomments{The complete version of this table is in the 
electronic edition of the Journal.  The printed edition contains 
only a sample.}
\end{deluxetable}